\documentclass[3p,authoryear, 11pt]{elsarticle}
\makeatletter
\def\ps@pprintTitle{%
	\let\@oddhead\@empty
	\let\@evenhead\@empty
	\def\@oddfoot{\centerline{\thepage}}%
	\let\@evenfoot\@oddfoot}
\makeatother
\usepackage{amssymb,amsmath}
\usepackage[hidelinks]{hyperref}
\usepackage[dvipsnames]{xcolor}
\usepackage{bm}

\usepackage{listings}
\usepackage{enumitem}
\usepackage{booktabs}

\usepackage{float}

\title{Spatiotemporal Multi-Resolution Approximations for Analyzing Global Environmental Data}

\author{Marius Appel\corref{cor1}\fnref{fn1}}%
\ead{marius.appel@uni-muenster.de}

\author{Edzer Pebesma}
\ead{edzer.pebesma@uni-muenster.de}

\address{Institute for Geoinformatics, University of M\"unster, Heisenbergstr. 2, 48149 M\"unster,
Germany}

\cortext[cor1]{Corresponding author}
\fntext[fn1]{Contact: marius.appel@uni-muenster.de}

\begin{document}

\begin{abstract}
Technological developments and open data policies have made large, global environmental datasets accessible to everyone. For analysing such datasets, including spatiotemporal correlations using traditional models based on Gaussian processes does not scale with data volume and requires strong assumptions about stationarity, separability, and distance measures of covariance functions that are often unrealistic for global data.
Only very few modeling approaches suitably model spatiotemporal correlations while addressing both computational scalability as well as flexible covariance models. In this paper, we provide an extension to the multi-resolution approximation (MRA) approach for spatiotemporal modeling of global datasets. MRA has been shown to be computationally scalable in distributed computing environments and allows for integrating arbitrary user-defined covariance functions. Our extension adds a spatiotemporal partitioning, and fitting of complex covariance models including nonstationarity with kernel convolutions and spherical distances. We evaluate the effect of the MRA parameters on estimation and spatiotemporal prediction using simulated data, where computation times reduced around two orders of magnitude with an increase of the root-mean-square prediction error of around five percent. This allows for trading off computation times against prediction errors, and we derive a practical strategy for selecting the MRA parameters. We demonstrate how the approach can be practically used for analyzing daily sea surface temperature and precipitation data on global scale and compare models with different complexities in the covariance function. 
\end{abstract}


\maketitle

\section{Introduction}
\label{introduction}

With the large amount of existing Earth observation programmes,
measurements of many environmental phenomena are available at global scale.
These include high-resolution optical imagery as well as atmospheric and
meteorological variables. At the same time, technological developments
such as Google Earth Engine \citep{Gorelick2017}, Earth observation data
cubes \citep{Mahecha2019, appel2019, Lewis2017}, and
array databases \citep{stonebraker2013, appel2018, baumann1998} make these data accessible to
scientists, such that studying changes, anomalies and interactions
between different phenomena becomes easier. In many cases, the data are
strongly autocorrelated in space and/or time and hence geostatistical models
are of particular interest to draw conclusions from the data. 

Continuous
phenomena are typically modeled as a Gaussian process (GP) defined by
a mean $\mu(\mathbf{s})$ and a covariance function 
$Cov(\mathbf{s}, \mathbf{s'})$ with
$\mathbf{s}, \mathbf{s'} \in D$ and typically
$D \subset \mathbb{R}^d$. Given a realization of a GP with
observations at $n$ locations $\mathbf{s}_1, ..., \mathbf{s}_n$, the
joint distribution is multivariate normal with mean vector
$\mathbf{\mu}_i = \mu(\mathbf{s}_i)$ and covariance matrix
$C_{ij} = Cov(\mathbf{s}_i,\mathbf{s}_j)$, $i,j = 1,..., n$. The
mean function typically integrates large scale trends or external
covariates whereas the covariance function determines the degree and
shape of spatial and/or temporal dependencies in the remainder process.
Since covariance matrices must be positive definite, a few commonly used
classes of covariance functions following a typical parameterization
including a spatial range, (partial) sill, and a nugget effect are mostly
applied. Inference typically concerns \emph{estimation} of the
mean and covariance function parameters and \emph{prediction} of measurements at
unobserved locations. Estimation and prediction are both affected by the big $n$
problem \citep{lasinio2013}, because the decomposition of the covariance matrix 
computationally scales with $O(n^3)$ and has memory
requirements $O(n^2)$. For parameter estimation, naive moments-based
experimental variogram estimation may lower the computational effort for simple stationary models. 

As a result, most practical applications of approaches based on decomposing the full covariance matrix are
limited to the order of $10^4 < n < 10^5$, whereas available global datasets such as
satellite-based precipitation observations from the Global Precipitation
Measurement (GPM) \citep{hou2014global} mission have more than $10^6$ observations 
per image, with new images being available every few hours or even more frequent. 
Furthermore, stationarity and separability assumptions in Geostatistical covariance
models are hardly valid on global scale, and using
Euclidean distances on projected coordinates or chordal distances on a sphere can be
disadvantageous compared to using spherical (great circle) distances
\citep{gneiting2013, banerjee2005, porcu2016}.

The big $n$ problem has been approached by a number of approximating
methods in the last years. Common methods aim at reducing the rank of
large covariance matrices \citep[see][for example]{cressie2008}, making covariance or their inverse precision
matrices sparse \citep[see][for example]{furrer2006, lindgren2011}, or partitioning the data into
smaller subsets \citep[see][for example]{katzfuss2017}. \citet{Heaton2019} give a comprehensive overview
of available approaches and provide a comparative benchmark for spatial
prediction, also including other algorithmic approaches such as gapfill
\citep{gerber2018}. The approaches considered have fundamentally
different underlying concepts and parameterizations. Resulting
differences in prediction performance as well as computation times still seem
remarkable. Each of the available approaches offer advantages and
disadvantages. Most important differences refer to the complexity of
supported covariance models (e.g.~nonstationarity), computational
behavior (e.g.~on distributed computing environments), whether the
domain can be extended to space \emph{and} time, and whether
approximations are valid on the sphere.

\sloppy
Concentrating on modeling global environmental phenomena as recorded 
from remote sensing satellites, only some of the approaches have been
successfully applied on large global spatiotemporal datasets. As such, 
Fixed Rank Kriging (FRK) \citep{cressie2008, zammitmangion2017} approximates a GP by a linear
combination of $r$ spatial basis functions with different resolutions.
$r$ is typically much smaller than $n$ and inference only needs to
factorize $r \times r$ matrices. The basis function representation
implies nonstationary covariances and is integrated in a spatial random
effects model to include external covariates in the mean function and
smaller scale variations and/or measurement errors.

Similarly, LatticeKrig \citep{nychka15} defines a multi-resolution representation 
where basis functions are compactly supported
and their coefficients are modeled as a Markov random field \citep{rue2005}. As a
result, computations make advantage of sparse matrix routines and a
larger number of basis functions that also capture small scale
variations can be used.

The stochastic partial differential equation (SPDE) approach \citep{lindgren2011} links Gaussian Markov Random
Fields and GPs with Mat\'{e}rn covariance function to achieve sparse
precision matrices by numerically solving stochastic partial
differential equation on a triangulated area of interest. The approach
has been applied in space and time \citep{cameletti2013}, is valid on
the sphere, and supports nonstationarity e.g.~by using covariates in the
dependence structure \citep{Ingebrigtsen2014}. Using integrated nested
Laplace approximations (INLA) \citep{rue2009}, the SPDE approach allows
for fast Bayesian inference.

Recently, \citet{zammitmangion2019} built on top of the SPDE approach
a multiscale representation that allows for working with large
datasets and nonstationary covariances. Results demonstrate that
approaches need to consider flexible covariance models as well as 
computational scalability.

In terms of scalable computing by making use of distributed computing
environments, especially the multi-resolution approximation (MRA) approach \citep{katzfuss2017} seems
promising. MRA recursively partitions the domain of interest into
smaller regions and assumes conditional independence between
observations in disjoint regions at the same partitioning level.
\citet{huang2019} provides a distributed implementation with an
application to remote sensing satellite data with $n > 10^7$.


In this paper, we use the multi-resolution approximation (MRA) approach
developed in \citet{katzfuss2017}, extend it to spatiotemporal domains
and covariance models, and evaluate the extent to which this can solve the
aforementioned difficulties with global environmental datasets. The
overall contribution of the paper is to make the MRA approach applicable
and available to global spatiotemporal datasets, and to evaluate and discuss its
performance and scalability, i.e., to what extent we can realistically
apply it on today's datasets.

The remainder of the paper is organized as follows. Section 2 describes
how we apply the MRA approach \citep{katzfuss2017} on spatiotemporal
data. Sections 3 and 4 present a simulation study and real world
examples, before Section 5 discusses the results, limitations, and gives an outlook on
future research directions. Section 6 concludes the paper.

\section{Multi-resolution Approximations in Space and Time}
\label{multi-resolution-approximations-in-space-and-time}

\begin{figure}[t]
\centering 
\includegraphics[width=11cm]{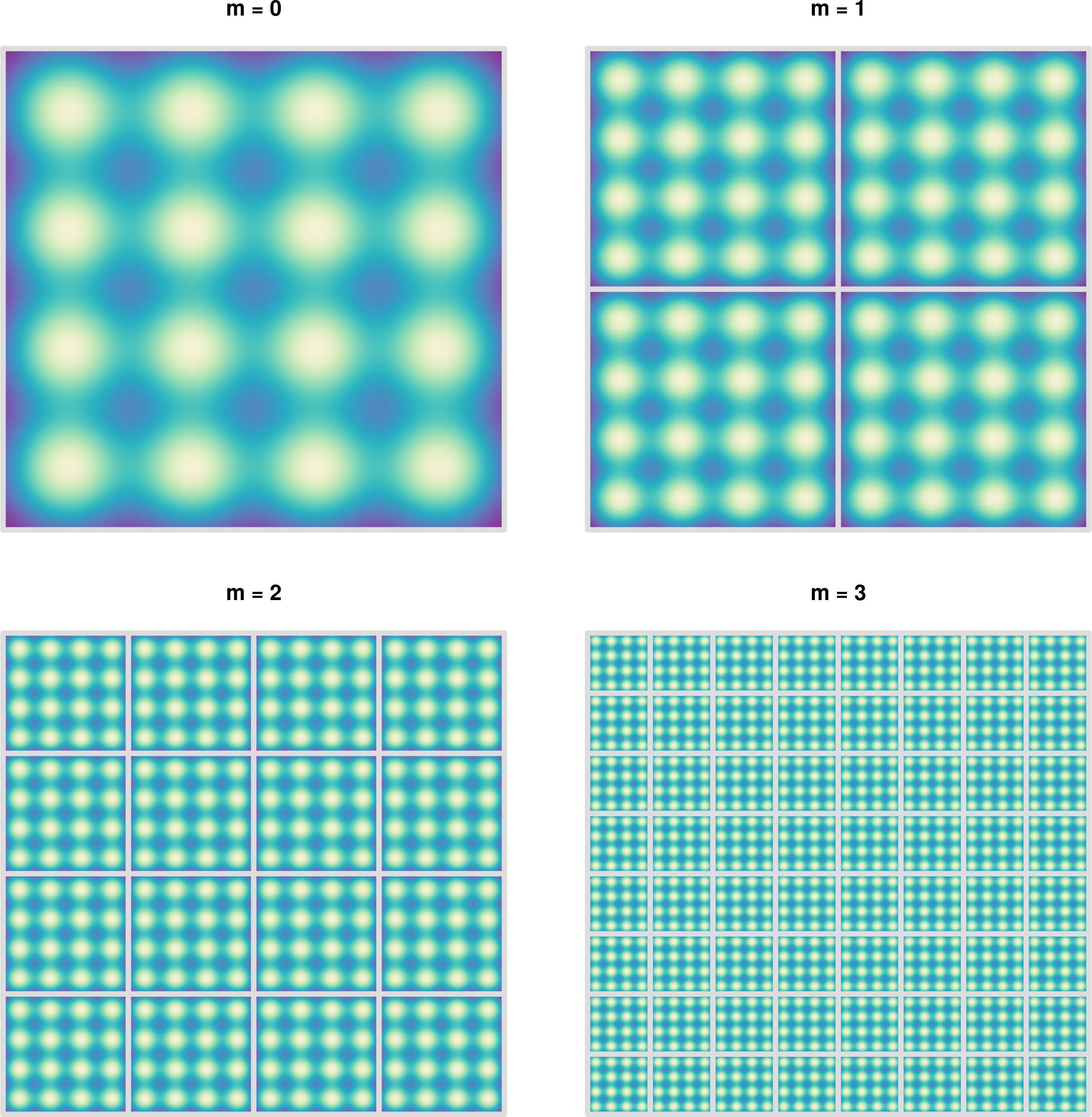} 
\caption[Illustration of the general MRA idea]{Illustration of the general MRA idea. Here, a spatial region is recursively partitioned into four smaller regions with $r=16$ basis functions; colors indicate the sum of these 16 functions.}
\label{fig:basis_functions}
\end{figure}

To approximate a mean zero Gaussian process, the MRA approach
\citep{katzfuss2017} recursively partitions the area of interest and
assumes conditional independence between different regions at the same
partitioning level $m \in \{0, ..., M\}$. Within each region, MRA defines a predictive process \citep{banerjee2008} at a set of $r$ knots, where larger 
regions (smaller $m$) capture large-scale spatial dependencies and smaller regions
consider the remainder process of the previous levels and hence capture
smaller scale spatial variations. As a result, approximations of covariances between any two locations are better if the locations share more regions. The basic idea is illustrated in Figure \ref{fig:basis_functions} with $M=3$ partitioning levels and $r=16$ (regularly placed) basis functions per region. Notice that for illustration purposes, the example uses identical basis functions at each level, although their shape is actually derived automatically with regard to a given covariance function. Regions in the lowest level may or may not correspond to actual observations. In general, the total number of basis functions is larger than the number of observations, making it
possible to account for both large and small scale variations.
Advantages in computational complexity are due to replacing the
decomposition of one big $n \times n$ matrix by the decomposition of
many smaller matrices. The partitioning into conditionally
independent regions allows for efficient inference on distributed
computing environments. Implementations of the approach in
\citet{huang2019} and \citet{jurek2018} have demonstrated computational
scalability up to $n \approx 10^7$. Below, we discuss how we use the
MRA approach to approximate spatiotemporal Gaussian processes. For
further details of the original approach and inference, the reader is
referred to \citet{katzfuss2017}.

\subsection{Spatiotemporal Partitioning}\label{spatiotemporal-partitioning}

Similar to the two-dimensional partitioning presented in
\citet{huang2019}, we apply a regular spatiotemporal partitioning, where
all three dimensions are divided into two equally sized parts per level.
A spatiotemporal region is split into $J=8$ smaller regions at each
level, resulting in $\sum_{i=0}^{M}8^i$ regions in total.  
Dimensions of spatiotemporal data can be rather uneven in size as
there are often more values along a spatial dimension than points in time. At the lower partitioning levels 
a regular partitioning would lead to regions with all observations recorded
at the same time. Due to the MRA assumption of conditional
independence between different regions at the same partitioning level,
this would prohibit estimations of temporal correlations.
For this reason, we maintain lower bound parameters
$(LB_x, LB_y, LB_t)$ specifying the minimum size of regions
allowed per dimension. For example, setting $LB_t = 10$ days makes sure that
the time dimension is not further partitioned if the temporal interval
size of regions would fall below 10 days. As a result, the number of
splits per region $J$ may vary by $m$ with $J \in \{2, 4, 8\}$
and the partitioning may stop at a higher level than a user may have
actually specified.

Given a number of basis functions per regions $r$ in the levels before
the last level ($m < M$), we first place
$\hat{r} = \lceil\sqrt[3]{r}\rceil^3$ knots at a regular spacetime
grid within each region from which we afterwards randomly sample exactly
$r$ knots. Knots refer to the locations where the basis functions of
the predictive process reach their maximum. To avoid duplicate knots in
a region and its child regions, which would lead to positive
definiteness issues, we assign a constant random spatiotemporal subpixel
shift to all knots in a region. A final check at the end of the
algorithm may furthermore remove duplicate knots in the unlikely event
that there are still duplicated knots. At the lowest partitioning level
($m = M$) we simply place knots at all available observation locations
within a region.

The overall partitioning algorithm is illustrated in pseudo code below.


\par\noindent\rule{\textwidth}{0.4pt}
\noindent\textbf{Input:}
	\begin{description}[itemindent=0mm, labelsep=2mm, leftmargin=5mm, topsep=2pt,itemsep=2pt,partopsep=0pt, parsep=0pt]
	\item[$extent$] \hfill \\
	Spatiotemporal extent of the region as lower and upper coordinates per
	dimension
	\item[$M$] \hfill \\
	Desired number of partitioning levels
	\item[$r$] \hfill \\
	Number of basis functions in regions at levels $m < M$
	\item[$LB$] \hfill \\
	Lower bounds of region sizes per dimension
	\item[$\mathbf{s}_i$, $i = 1,..., n$] \hfill \\
	Observation locations
	\item[$z_i$, $i = 1,..., n$] \hfill \\
	Observation values
	\item[$\hat{\mathbf{s}}_j$, $j = 1,..., m$] \hfill \\
	Optional locations for prediction
	\end{description}

\noindent\textbf{function} partition\_region($extent$, $m=0$)
\begin{enumerate}[itemindent=0mm,labelsep=2mm, topsep=0pt,itemsep=0pt,partopsep=0pt,parsep=0pt,label={\arabic*.}]
\item Place $r$ knots within the current region:
  \begin{enumerate}[itemindent=0mm, labelsep=2mm, leftmargin=3mm, topsep=0pt,itemsep=0pt,partopsep=0pt,parsep=0pt,label={}]
 \item
    \textbf{if} ($m = M$)
    \begin{enumerate}[itemindent=0mm, labelsep=2mm, leftmargin=8mm, topsep=0pt,itemsep=0pt,partopsep=0pt,parsep=0pt,label={\arabic*.}]
    \item Find observation locations $\mathbf{s}_i$ which lie within extent
    \item Add found knots and their data values $z_i$ to the current region
    \item \textbf{if} available, add prediction locations  $\hat{\mathbf{s}}_j$ lying within extent to the current region
    \end{enumerate}
  \item \textbf{else}
    \begin{enumerate}[itemindent=0mm, labelsep=2mm, leftmargin=8mm, topsep=0pt,itemsep=0pt,partopsep=0pt,parsep=0pt,label={\arabic*.}]
    \item Create a regular grid at $\hat{r} = \lceil\sqrt[3]{r}\rceil^3$,
      knots ($\sqrt[3]{\hat{r}}$ per dimension) within the extent
    \item Apply a small random sub-pixel shift per dimension to avoid
      identical knot locations at different $m$
    \item Randomly sample exactly $r$ of the $\hat{r}$ knots
    \end{enumerate}
   \item \textbf{end if}
  \end{enumerate}
\item Add region to the list of output regions
\item Partition the current region:
\begin{enumerate}[itemindent=0mm, labelsep=2mm, leftmargin=3mm, topsep=0pt,itemsep=0pt,partopsep=0pt,parsep=0pt,label={}]
\item \textbf{if ($m < M$)}
  \begin{enumerate}[itemindent=0mm, labelsep=2mm, leftmargin=8mm, topsep=0pt,itemsep=0pt,partopsep=0pt,parsep=0pt,label={\arabic*.}]
  \item Split all dimensions into two equally sized intervals if size of the
    current region is larger than 2 times the corresponding $LB$
  \item \textbf{if} no dimension has been splitted: \textbf{exit}
  \item Combine all dimension intervals as $J=2,4,$ or $8$
    spatiotemporal regions with extents
    $extent.sub[i=1,...,J]$
  \item
    \textbf{for} all $i \in 1,...,J$ regions, call \\ partition\_region($extent.sub[i]$,  $m+1$)
  \end{enumerate}
\item \textbf{end if}
\end{enumerate}
\end{enumerate}
\noindent\textbf{end function}
\par\noindent\rule{\textwidth}{0.4pt}

The recursive function starts with $m=0$ and produces a list (containing a tree structure) of regions, their spatiotemporal extents, knot locations, and at the lowest level the observations. For spatiotemporal prediction, the corresponding locations are furthermore added as knots to regions in the lowest level. 

\subsection{Spatiotemporal Covariance Functions}
\label{spatiotemporal-covariance-functions}

While the selection of covariance models in space mostly concerns a few predefined
functions, the selection of appropriate models in space and time is
generally more difficult. A straightforward approach referred to as the
\emph{metric model} is to interpret spacetime as three-dimensional
Euclidean space, scale the temporal coordinates by an anisotropy factor
and use the same covariance functions as for purely spatial models.
Other simple models may be simple products and sums of purely spatial,
purely temporal, and metric models \citep[see][for a
comprehensive discussion of frequently used models and
parameterizations]{graeler2016}. For example the \emph{separable} model is a simple
product of a purely spatial and a purely temporal covariance functions
with computational advantages in the decomposition of their
covariance matrices using the Kronecker product.

One advantage of the MRA approach is that covariance functions can be
completely defined by users. As such, users simply need to implement a
function that creates a valid covariance matrix from two given sets of
locations. As a result, simple models can be directly integrated into
the MRA approach without any additional work. On a global scale,
however, these models make unrealistic assumptions (see Section 1).

To make the approach applicable on spherical distances and add
nonseparability, we have integrated some of the covariance functions suggested in \citet{porcu2016}. To incorporate nonstationarity,
we have integrated a kernel convolution approach \citep{stein2005,
Risser2015,paciorek2006spatial}, 
where the variance, smoothness, and anisotropy of the covariance model vary spatially.
In our implementation (Section \ref{implementation-in-r}), we represent these processes as
a linear combination of Gaussian radial basis functions centered at a
few spatiotemporal locations. As an alternative, it would be possible to
apply space and/or time deformations \citep{sampson1992} to yield
nonstationary covariances in the MRA approach.

\subsection{Parameter Estimation and Prediction}
\label{parameter-estimation-and-prediction}

We use the original MRA formulation for likelihood-based inference from
\citet{katzfuss2017} and omit details here. To estimate non-negative 
parameters of covariance functions, one can use bound constrained optimization algorithms such as
L-BFGS-B \citep{byrd1995} or BOBYQA \citep{powell2009}, or use general
purpose optimization algorithms to estimate parameters on logarithmic
scale.

Since prediction with MRA is known to produce artefacts at region
boundaries, we furthermore implement a simple averaged prediction using multiple
partitions. Splits in each dimension can be shifted by a fraction
of the size of regions at the lowest partitioning level towards lower and upper values, yielding 9 different partitions. Computation times increase accordingly.

\subsection{Implementation in R}
\label{implementation-in-r}

The spatiotemporal MRA approach has been implemented prototypically as an
add-on package\footnote{\url{https://github.com/appelmar/stmra}} to
the R language and environment for statistical computing
\citep{rcore2020}. The package includes functions for parameter
estimation and prediction and includes the covariance
functions described above. It interfaces R packages \texttt{raster}
\citep{Hijmans2019} and \texttt{gdalcubes} \citep{appel2019} to read
satellite time series data. Below, a simple R script illustrates the
usage of the package for parameter estimation and prediction of sea surface temperature data. 

\begin{lstlisting}[language=R,stringstyle=\footnotesize\ttfamily,basicstyle=\footnotesize\ttfamily,showstringspaces=false,frame=lines,xleftmargin=0cm,belowskip=10pt]
library(stmra)
sst = raster::stack(
  list.files(system.file("sst", package="stmra"),
    pattern = ".tif", full.names = TRUE)[1:10])

sst = sst - mean(as.vector(sst), na.rm =TRUE)
part = stmra_partition(M = 4, r = 16, domain = sst, 
  region_minsize = c(0,0,5))

# pars: sill, spatial range, temporal range,
#       spatial nugget, temporal nugget
lo = rep(1e-4, 5)
up = rep(1e5, 5)
theta0 = c(2, 10, 5, 0.1, 0.1)

# parameter estimation (use 10000 observations)
v = as.vector(sst)
v[-sample(which(!is.na(v)),10000)] = NA
training_data = raster::setValues(sst, v)
model = stmra(part, stmra_cov_separable_exp, 
  training_data, theta0 = theta0, lower_bounds = lo,
  upper_bounds = up, trace = TRUE, 
  control=list(ftol_abs = 1, maxeval = 200))

# prediction (without land areas)
mask_spatial = raster::mean(sst, na.rm = TRUE)
mask_spacetime = sst
for (i in 1:raster::nlayers(mask_spacetime)) {
  mask_spacetime[[i]] = mask_spatial
}
pred = predict(model, data = training_data, 
  mask = mask_spacetime)

plot(pred)
plot(pred, variance = TRUE)
\end{lstlisting}

The implementation is serial only and uses the
original implementation from the supplementary materials of
\citet{katzfuss2017} for inference. \citet{huang2019} provide a
multithreaded and distributed C++ implementation using OpenMP
\citep{dagum1998} and Message Passing interface (MPI) \citep{mpi2012},
where even the serial implementation outperforms a compared MATLAB
implementation. A similar implementation interfacing C++ and R via the
\texttt{Rcpp} package \citep{Eddelbuettel2011} is currently work in progress.

\section{Simulation Study}
\label{simulation-study}

\begin{table}[t]

\caption[Initial values and bounds for parameter estimation in the simulation study]{\label{tab:sim_pars}Initial values and bounds for estimating spatial range ($\rho$), partial sill ($\sigma^2$), nugget effect ($\tau^2$), and spatiotemporal anisotropy ($s$). Initial values of $\sigma^2$ and $\tau^2$ refer to 1 and 0.1 times the empirical variance of the data.}
\centering
\begin{tabular}{lrrrr}
\toprule
  & $\rho$ & $\sigma^2$ & $\tau^2$ & $s$\\
\midrule
Initial Value & 0.5 & 0.875 & 0.088 & 1\\
Lower Bound & 0.001 & 0.001 & 0.001 & 0.001\\
Upper Bound & 1 & 3.502 & 3.502 & 50\\
\bottomrule
\end{tabular}
\end{table}

We apply the described spacetime MRA approach on a simulated dataset and
evaluate how the MRA parameters affect quality of predictions and
parameter estimations as well as computation times. Using the R package
\texttt{RandomFields} \citep{schlather2015}, we simulated a three-dimensional regular spacetime grid with size
$n = 50 \times 50 \times 50 = 125000$, following a metric
spatiotemporal model with an exponential covariance
function (range = 0.2, partial sill = 0.05, spatiotemporal anisotropy =
0.02, nugget = 0.05). Figure \ref{fig:simdata} shows the first 20 time
slices of the dataset.

To assess the performance of the spatiotemporal MRA, we perform
parameter estimation for different values of $M$ and $r$ and compare
obtained values to the true values. For assessing prediction
performance, we assume the true parameters to be known and predict 90\%
randomly selected pixels again for different values of $M$, and $r$.
We calculate several diagnostic scores, including root-mean-square
prediction error (RMSE), mean absolute error (MAE), the fraction of true
values lying within a $\mu \pm 2\sigma$ prediction interval (COV2SD),
and measure computation times. Predictions for the same parameter values
have been repeated three times each. Due to the randomness in the
selection and shifting of knots (see Section 2), results are not exactly identical.
To find the parameters with maximum MRA likelihood, we apply the
L-BFGS-B optimization algorithm with starting values and bounds as presented in Table \ref{tab:sim_pars}.

\subsection{Parameter Estimation Results}
\label{parameter-estimation-results}

\begin{figure}
\centering 
\includegraphics[width=11cm]{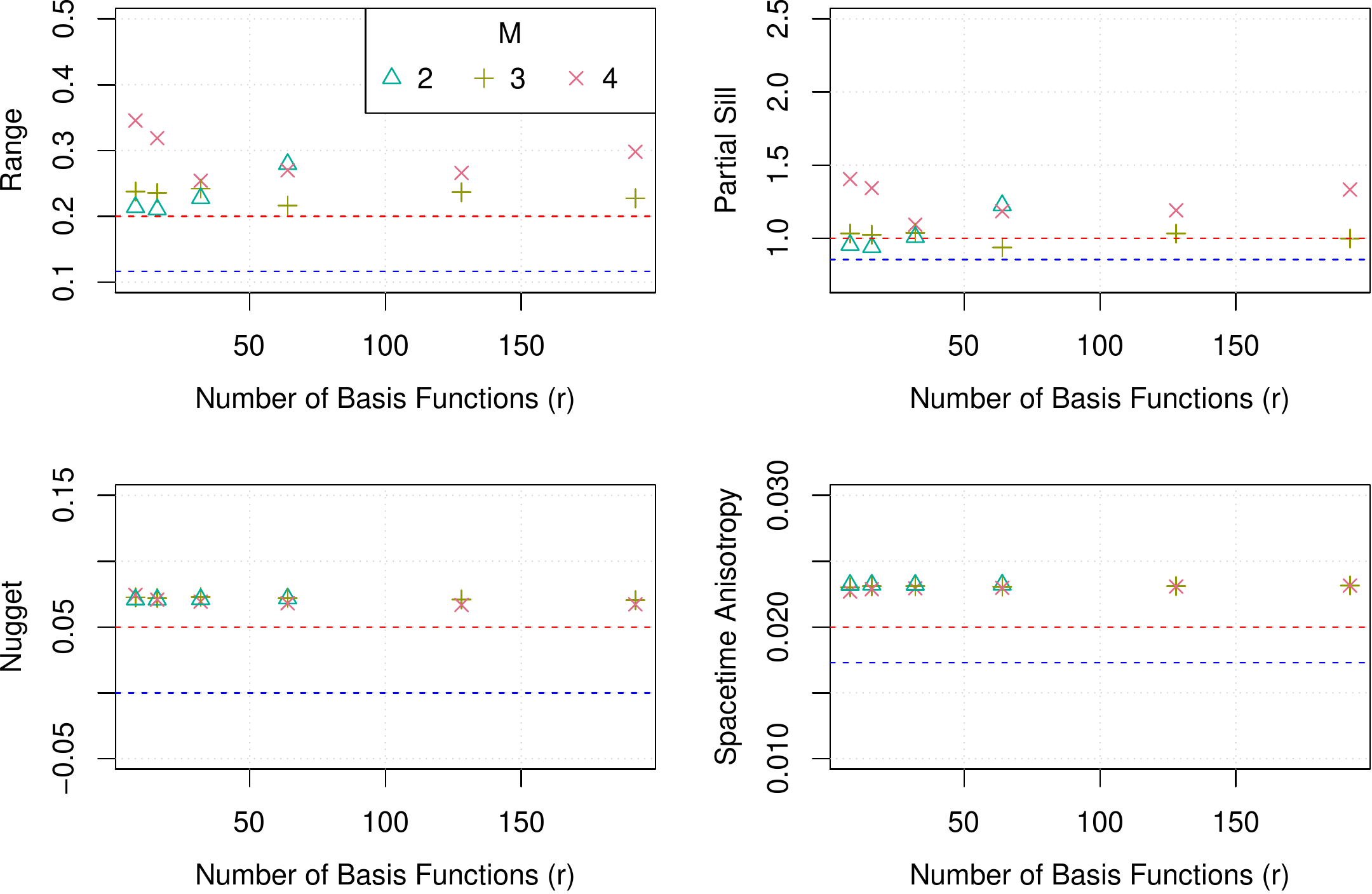} 
\caption[Parameter estimation results of the simulation study]{\label{fig:fig_results_estimation_sim}Parameter estimation results of the simulation study. The horizontal lines represent the true value (dashed red) and the best fit based on an experimental variogram after trying different starting values for numerical optimization (dashed blue). $M$ refers to the number of partitioning levels. Due to long computation times, only up to $r=64$ basis functions have been used for $M=2$. }
\label{fig:sim_estimation_results}
\end{figure}

Figure \ref{fig:fig_results_estimation_sim} shows parameter estimates
against the true values. The worst estimates have been obtained for
$M=4$ and $r=8$. In general, the quality of estimations seems relatively independent of both $M$ and
$r$, although for $M=4$, estimates of the spatial range and partial sill tend to be slightly worse for low $r$.

We have also performed traditional experimental variogram estimation
using the \texttt{gstat} R package
\citep{graeler2016, pebesma2004}. After using a large number of
different initial parameter values (including the true values) for
fitting the theoretical variogram, the blue
horizontal dashed lines in Figure \ref{fig:fig_results_estimation_sim}
show the estimates closest to the true values. Most MRA estimations are closer or at similar distance to the true values than the best solution of experimental variogram fitting, which also produced a large number of much worse estimates. For example, the median variogram estimate 
is 0.32 for the nugget effect and 0.46 for the spatial range, i.e., worse than the worst MRA estimates.
The likelihood-based MRA fitting seems more robust and traditional variogram fitting furthermore may be impracticable for more complex nonstationary covariance models.

In terms of computation times, the MRA approach is naturally much slower
than traditional variogram estimation. Single evaluations of the MRA likelihood took between 11 seconds ($M=4, r=8$) and 40 minutes ($M=2, r=32$), where the number of likelihood evaluations ranged from 95 to 864 until convergence. For more complex models with many more parameters, even more iterations might be needed.

\subsection{Prediction Results}
\label{prediction-results}

\begin{figure}
\centering 
\includegraphics[width=11cm]{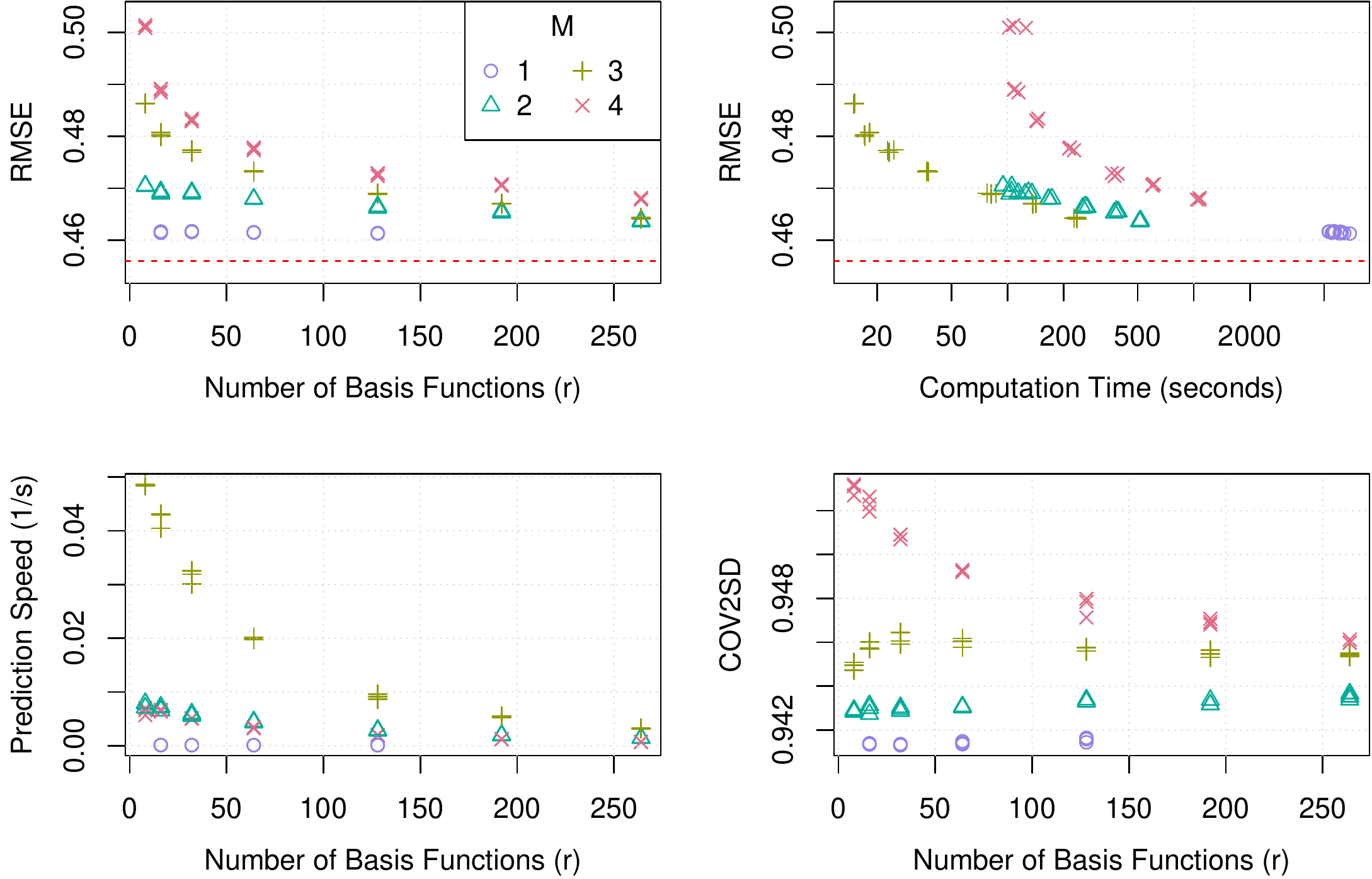} 
\caption[Prediction results of the simulated dataset]{\label{fig:fig_results_prediction_sim}Prediction results of the simulated dataset showing root-mean-square prediction error (RMSE) of prediction times against the number of basis functions $r$ per region in partitioning levels $m < M$ (top left),  RMSE against computation times (top right), prediction speed computed as $R^2$ divided by computation time, quantifying the time-efficiency of predictions (bottom left), and the fraction of the true values lying within a $\mu \pm 2 \sigma$ prediction interval (COV2SD, bottom right). Horizontal dashed red lines refer to Kriging prediction.  Due to long computation times, only up to $r=128$ basis functions have been used for $M=1$.}\label{fig:sim_prediction_results}
\end{figure}

Figure \ref{fig:fig_results_prediction_sim} shows diagnostic scores
of the predictions. Computation times ranged from approximately 15
seconds ($M=3, r=8$) to approximately 114 minutes ($M=1, r=128$)
where RMSE and MAE improved from 0.486 to 0.461 and 0.387 to 0.367
respectively. The number of basis functions thereby clearly affects both
computation times, and prediction performance. The effect of $r$ on
computation times and prediction performance is less for smaller values
of $M$, because regions at the lowest level are larger and hence
contain more observations. For the example of $M=1$, regions in the
lowest level $m=1$ contain 15625 observations compared to only a very
low number of $r$ basis functions at $m=0$.

Interestingly, it was possible to achieve similar prediction errrors
with $M=3$ and $M=2$ when using more basis functions for $M=3$
with very similar computation times too. Figure
\ref{fig:fig_results_prediction_sim} furthermore shows achieved
$R^2$ values normalized by computation times as a simple indicator for
the speed or efficiency of predictions, where $M=3$ is generally
fastest. This suggests a practical strategy how to select the MRA
parameters by finding the most efficient $M$ first (e.g.~by starting
at the $M$ where regions in the lowest level contain $\approx$ 100
observations), and increasing $r$ as computation times allow
afterwards.

Traditional Kriging interpolation using the R package \texttt{gstat}
\citep{pebesma2004, graeler2016} yield RMSE values that were
1.17\% better than the best MRA result ($M=1, r=128$) and 6.64\%
better than the fastest ($M=3, r=8$) result, with computation times 
around 12 hours.

In all cases, MRA predictions provide relative good estimations of the
variances, such that around 95.45\% of the true values lie within a
$\mu \pm 2 \sigma$ prediction interval. As $r$ increases these
values tend to be more similar even for different values of $M$.

\section{Applications to Real World Data}
\label{applications-to-real-world-data}

Below, we demonstrate how the spacetime MRA approach can be applied on two
global spatiotemporal datasets and how more complex nonstationary
covariance models with nonstationarity and spherical (Great circle) distances 
can be integrated.

\subsection{Daily Sea Surface Temperatures}
\label{daily-sea-surface-temperatures}

\begin{figure}
\centering 
\includegraphics[width=11cm]{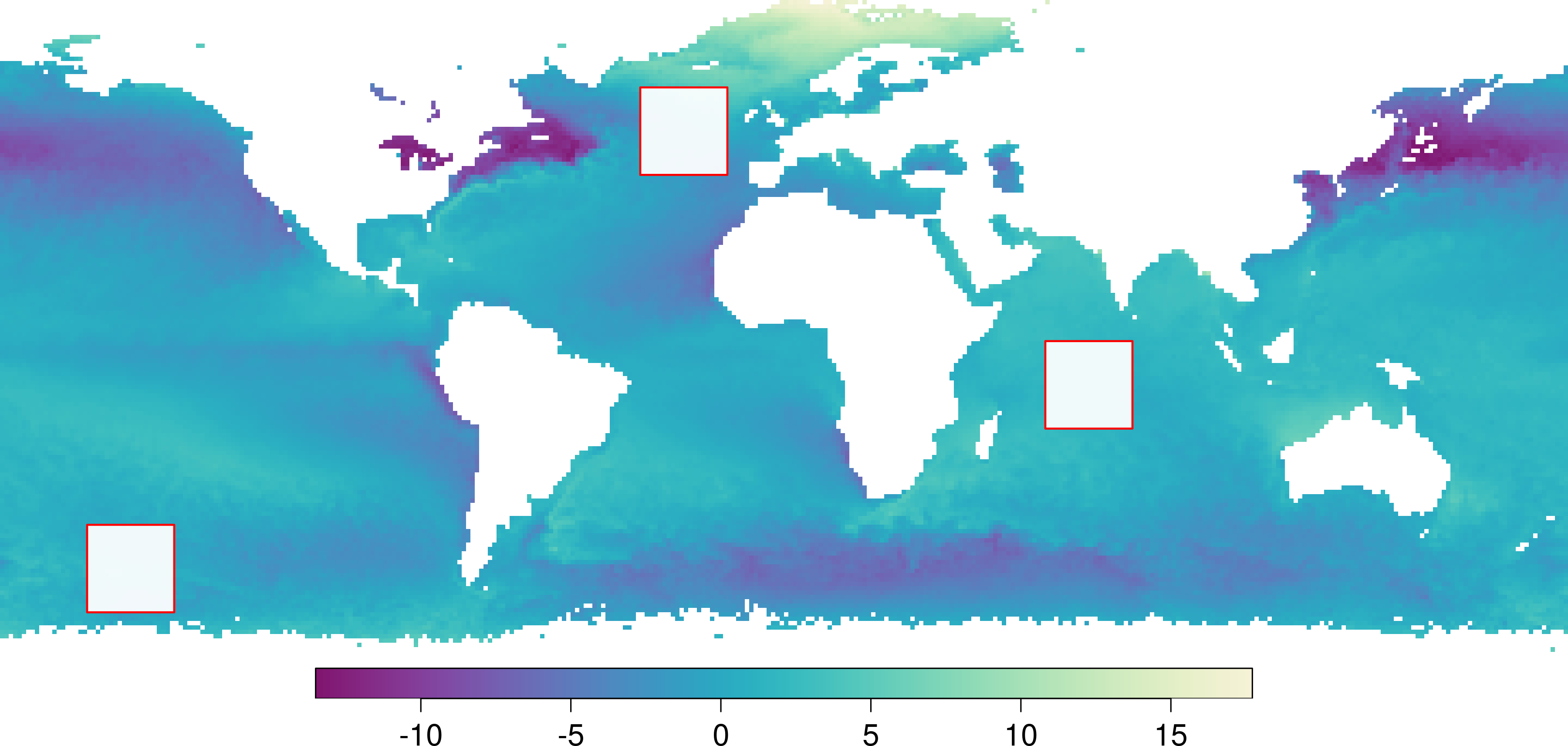} 
\caption[Temporal mean values of sea surface temperature residuals after fitting a simple quadratic polynomial over latitude]{Temporal mean values of sea surface temperature residuals after fitting a simple quadratic polynomial over latitude. White boxes refer to the three regions used for block validation.}
\label{fig:sst_blockvalidation}
\end{figure}

We apply the MRA approach on global daily sea surface temperature (SST)
measurements from the Moderate-resolution Imaging Spectroradiometer
(MODIS)\footnote{see \url{https://doi.org/10.5067/MODST-1D9D4} (accessed
  on 2020-02-19)}, recorded between 2018-04-10 and 2018-05-09. To keep
reasonable computation times in this example, we averaged original
pixels (with approximate size of 9km by 9km) to a $1^{\circ} \times 1^{\circ}$ grid using the
geospatial data abstraction library (GDAL) \citep{warmerdam2008}. The
resulting dataset contains gaps and sums to $n=637770$ valid
observations. Following \citet{zammitmangion2019}, we first fit a
quadratic polynomial over latitude and apply MRA models on the
residuals. Figure \ref{fig:sst_blockvalidation} illustrates mean residuals of the models over pixel time series and Figure \ref{fig:sst_data_all} in the Appendix shows the full dataset. After some initial experimentation, we choose $M=4$, vary the number of basis functions, and apply the following three different covariance models, where $\bm{\theta}$ is a vector of parameters, $\Delta_s, \Delta_t$ represent spatial and temporal distances between any two locations (either Euclidean or spherical), and
$\mathbf{1}(\cdot)$ is the indicator function:

\begin{enumerate}[label={\texttt{M\arabic*}:}]
\item A stationary separable spatiotemporal model with two
  exponential covariance functions and a joint sill ($\bm{\theta}_1$), separate ranges ($\bm{\theta}_2, \bm{\theta}_3$) and
  nugget parameters ($\bm{\theta}_4, \bm{\theta}_5$), using naive spatial Euclidean distances of latitude
  and longitude coordinates: { \small \begin{align*} 
  \textrm{cov}(\Delta_s,\Delta_t) =& ~ \bm{\theta}_1 \textrm{cov}_s(\Delta_s) ~ \textrm{cov}_t(\Delta_t) \\
  \textrm{cov}_s(\Delta_s) =& ~(1 - \bm{\theta}_4 + \bm{\theta}_4 \mathbf{1}(\Delta_s = 0))  \exp(-\Delta_s/\bm{\theta}_2) \\ 
  \textrm{cov}_t(\Delta_t) =& ~(1 - \bm{\theta}_5 + \bm{\theta}_5 \mathbf{1}(\Delta_t = 0)) \exp(-\Delta_t/\bm{\theta}_3) \\
  \end{align*}}
\item A stationary nonseparable covariance model with spherical
  (Great circle) distances from \citet[][Equation 15]{porcu2016}, with partial sill ($\bm{\theta}_1$), nugget ($\bm{\theta}_4$), and spatial ($\bm{\theta}_2$) and temporal ($\bm{\theta}_3$) scaling parameters:
  {\small \begin{align*} 
  \textrm{cov}(\Delta_s, \Delta_t) = \frac{\bm{\theta}_1 + \bm{\theta}_4 \mathbf{1}(\Delta_s = \Delta_t = 0)}{1 + \frac{\Delta_s}{\bm{\theta}_2}} \mathrm{exp}\left(- \frac{\Delta_t}{\bm{\theta}_3 \left(1 + \frac{\Delta_s}{\bm{\theta}_2}\right)^{1/4} } \right)
  \end{align*}}
\item A model similar to \texttt{M1} that introduces
  nonstationarity in the spatial covariance function using the kernel
  convolution approach \citep{paciorek2006spatial, Risser2015},
  with standard deviation and geometric anisotropy (defined by
  length of the east-west and south-north axes) processes varying by latitude as a
  linear combination of Gaussian radial basis functions centered at
  (pseudo) latitudes $\pm 110^{\circ}$, $\pm 82.5^{\circ}$,
  $\pm 55^{\circ}$, $\pm 27.5^{\circ}$, and $0^{\circ}$. Let
  $\mathbf{s}_1 = (\lambda_1, \phi_1, t_1)^T$ and
  $\mathbf{s}_2 = (\lambda_2, \phi_2, t_2)$ represent latitude,
  longitude, and time coordinates of two points. The model
  can be defined as:  {\small \begin{align*} 
  \textrm{cov}(\mathbf{s}_1, \mathbf{s}_2) =& ~ \bm{\theta}_1 \textrm{cov}_s(\mathbf{s}_1, \mathbf{s}_2) ~ \textrm{cov}_t(|t_1 - t_2|) \\
  \textrm{cov}_t(\Delta_t) =& ~(1 - \bm{\theta}_2 + \bm{\theta}_2 \mathbf{1}(\Delta_t = 0)) \exp(-\Delta_t/\bm{\theta}_3) \\
  \textrm{cov}_s(\mathbf{s}_1, \mathbf{s}_2) =& ~(1 - \bm{\theta}_1 + \bm{\theta}_1 \mathbf{1}(\mathbf{s}_1 = \mathbf{s}_2)) C(\mathbf{s}_1, \mathbf{s}_2) \\ 
  C(\mathbf{s}_1, \mathbf{s}_2) =& ~\sigma(\lambda_1) \sigma(\lambda_2) \left| \frac{\mathbf{\Sigma}(\lambda_1) + \mathbf{\Sigma}(\lambda_2)}{2} \right|^{-1/2} \exp(-\sqrt{Q}) \\
  Q =& ~(\mathbf{s}_1 - \mathbf{s}_2)^T \left( \frac{\mathbf{\Sigma}(\lambda_1) + \mathbf{\Sigma}(\lambda_2)}{2} \right)^{-1} (\mathbf{s}_1 - \mathbf{s}_2) \\
  \sigma(\lambda) =&  ~RB(\lambda, \bm{\theta}_{4,...,12}) \\
  \mathbf{\Sigma}(\lambda) =&  ~\begin{pmatrix}
  RB(\lambda, \bm{\theta}_{13,...,21}) & 0 \\
  0 & RB(\lambda, \bm{\theta}_{22,...,30}) 
  \end{pmatrix} \\
  RB(\lambda, \mathbf{\omega}) =& ~\sum_{i = 1}^{9} \omega_i  \exp\left(-(|\lambda - \mathbf{k}_i | / 20)^2\right) \\
  \mathbf{k} =& ~(-110, -82.5, -55, -27.5, 0, 27.5, 55, 82.5, 110) \\
  \end{align*}}
\end{enumerate}

For all models, parameters are first estimated using the same 100000 randomly selected observations before we use two different strategies to validate prediction performance. First, we use the same 100000 observations to predict the remaining data and, at the same time, produce gap-free image time series. We refer to this validation strategy as \emph{random validation}. Second, we leave
out and predict three spatiotemporal regions of size $20^{\circ} \times 20^{\circ} \times 9$ days
in the lower, middle and higher latitudes to assess large scale
prediction performance (referred to as \emph{block validation}, see Figure \ref{fig:sst_blockvalidation}).
We calculate the root-mean-square prediction error (RMSE), mean absolute error (MAE), and count how many of the true
observations lie within a $\mu \pm 2\sigma$ prediction interval (COV2SD).

\begin{figure}
	\centering 
	\includegraphics[width=5.4cm]{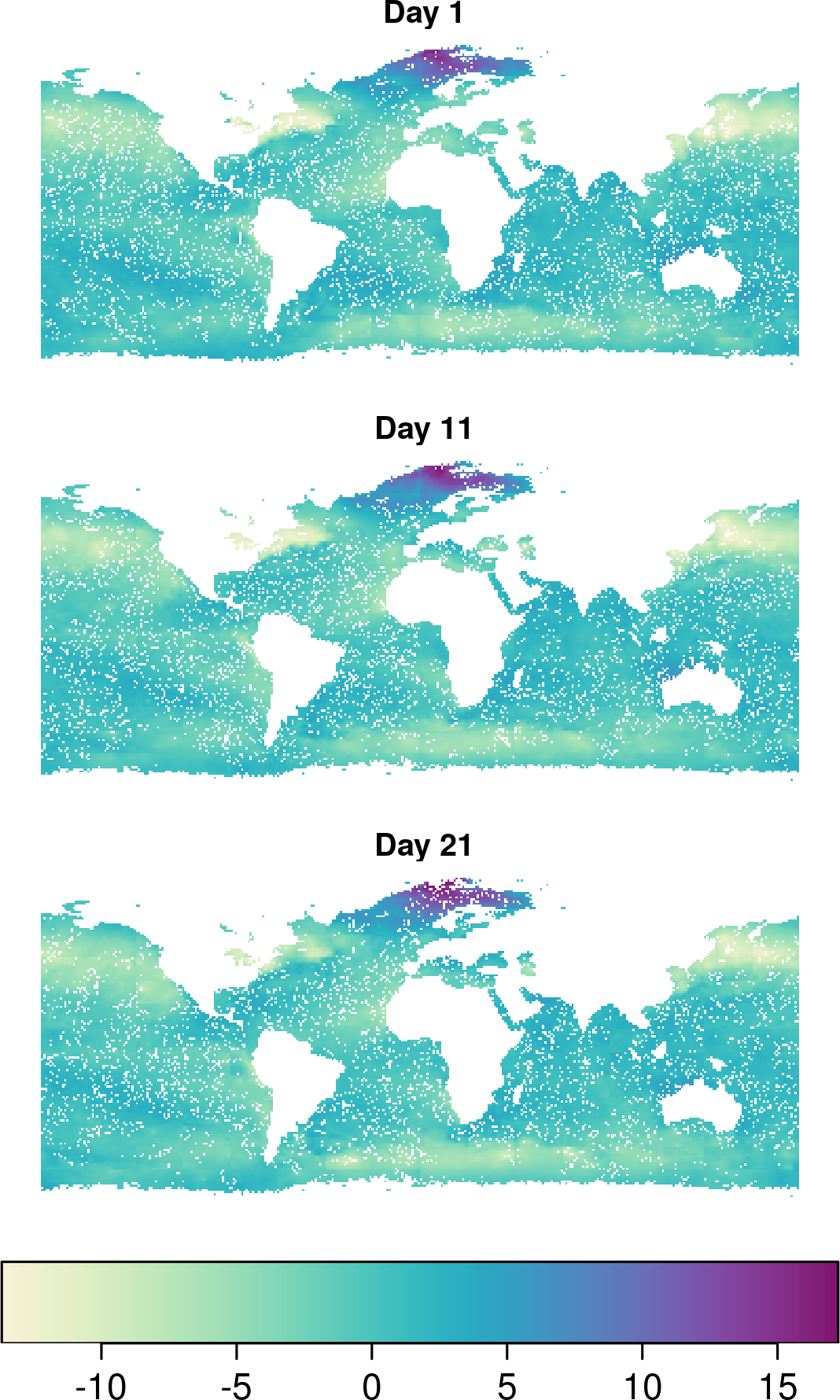} \includegraphics[width=5.4cm]{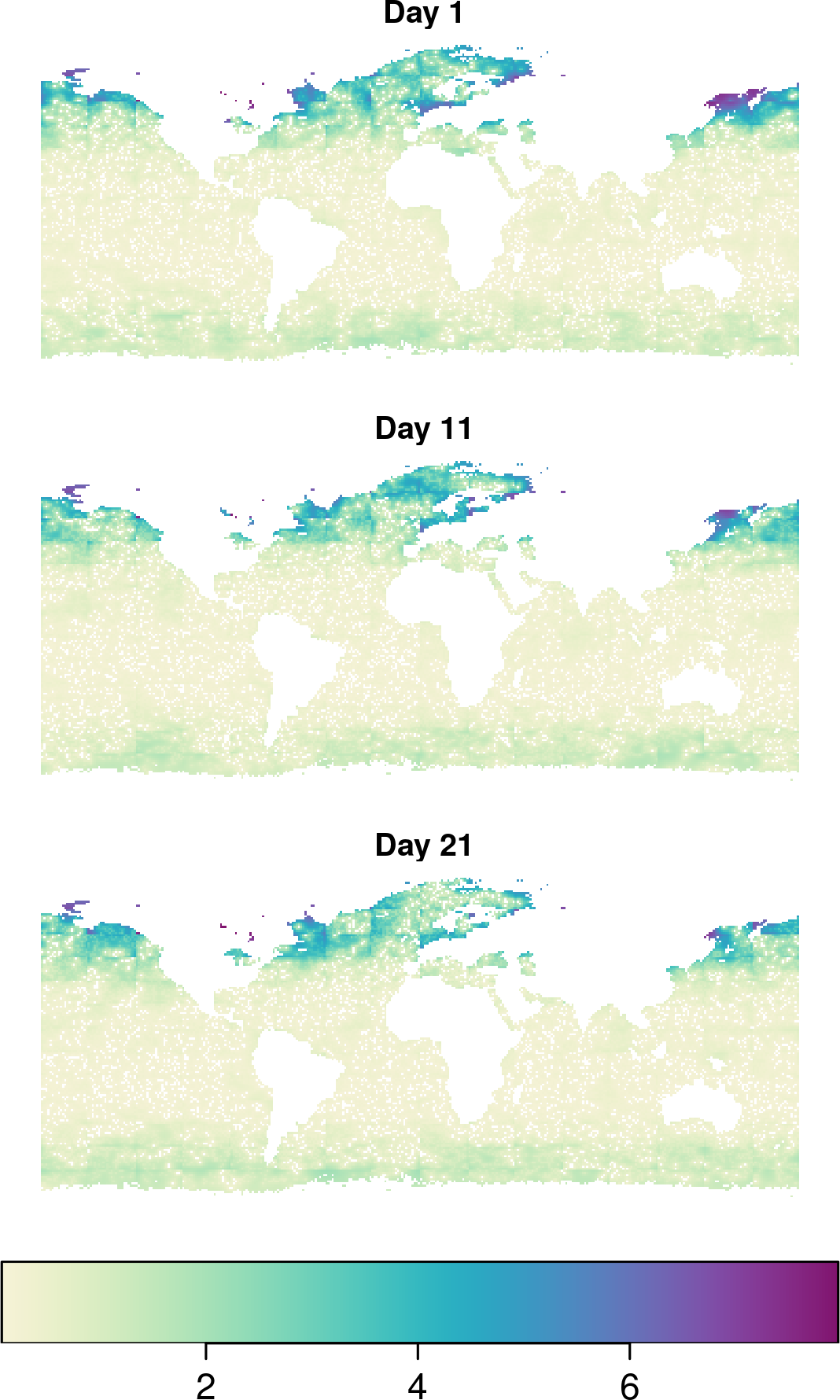} 
	\caption[Predicted mean and variance for model M3 at $M=4, r=32$ at three different days]{\label{fig:fig_sst_prediction}Predicted mean (left) and variance (right) for model M3 at $M=4, r=32$ at three different days. }\label{fig:sst_predictions}
\end{figure}

\begin{figure}
\centering 
\includegraphics[width=11cm]{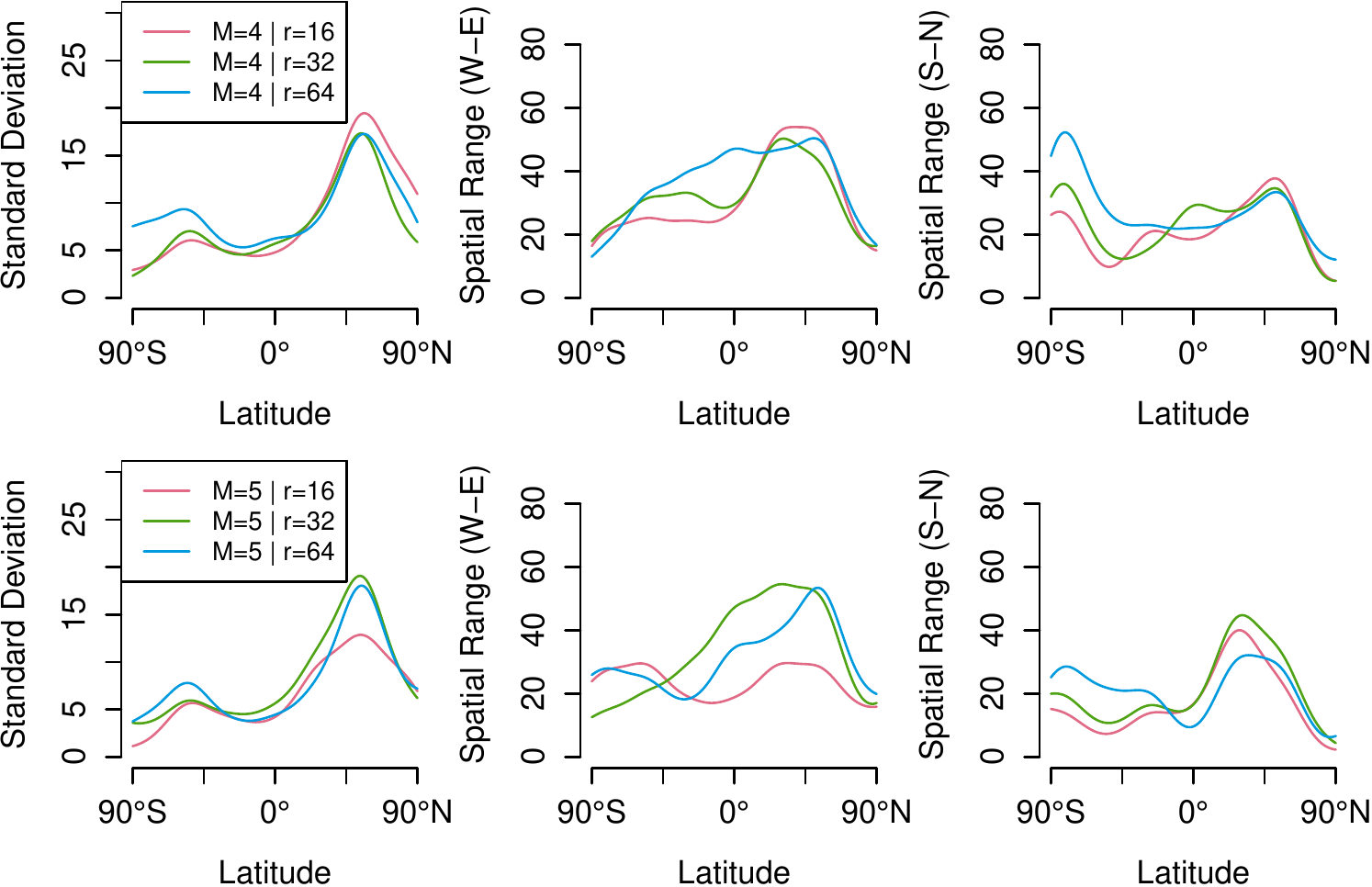} 
\caption[Estimates of spatially varying standard deviation and anisotropy processes of model \texttt{M3} in the sea surface temperature data example]{Estimates of spatially varying standard deviation and anisotropy processes of model \texttt{M3} in the sea surface temperature data example.}\label{fig:sst_M3pars}
\end{figure}

\begin{table}[t]
\caption[Results of predictions in the sea surface temperature data example]{\label{tab:sst_results}Results of predictions in the sea surface temperature data example.}
\centering
\scriptsize
\setlength{\tabcolsep}{0.8mm}
\begin{tabular}{lllrrrrrrrr}
\toprule
\multicolumn{3}{c}{ } & \multicolumn{4}{c}{Random Validation} & \multicolumn{4}{c}{Block Validation} \\
\cmidrule(l{3pt}r{3pt}){4-7} \cmidrule(l{3pt}r{3pt}){8-11}
Model & M & r & RMSE & MAE & COV2SD & $t$ (min) & RMSE & MAE & COV2SD & $t$ (min)\\
\midrule
M1 & 4 & 16 & 0.694 & 0.454 & 0.940 & \textbf{20} & 0.831 & 0.577 & 0.944 & 76\\
M1 & 4 & 32 & 0.694 & 0.454 & 0.940 & 26 & 0.839 & 0.585 & 0.937 & \textbf{72}\\
M1 & 4 & 64 & 0.693 & 0.454 & 0.941 & 35 & 0.842 & 0.585 & 0.937 & 80\\
\addlinespace[0.2em]
M2 & 4 & 16 & 0.694 & 0.454 & 0.940 & 31 & 0.854 & 0.592 & 0.938 & 78\\
M2 & 4 & 32 & 0.695 & 0.454 & 0.940 & 34 & 0.863 & 0.599 & 0.938 & 78\\
M2 & 4 & 64 & 0.694 & 0.454 & 0.939 & 44 & 0.868 & 0.603 & 0.936 & 82\\
\addlinespace[0.2em]
M3 & 4 & 16 & \textbf{0.690} & \textbf{0.447} & \textbf{0.948}& 141 & \textbf{0.781} & \textbf{0.543} & 0.968 & 125\\
M3 & 4 & 32 & 0.694 & 0.448 & 0.945 & 151 & 0.787 & 0.548 & \textbf{0.962} & 132\\
M3 & 4 & 64 & 0.691 & 0.448 & 0.943 & 171 & 0.802 & 0.557 & \textbf{0.962} & 144\\
\bottomrule
\end{tabular}
\end{table}
Table \ref{tab:sst_results} shows the obtained prediction scores and
computation times. In general, the nonstationary model (\texttt{M3})
performs best in terms of RMSE and MAE. Differences are stronger for block
validation where larger missing regions are predicted. The
nonseparable spherical model and the simple separable model present very
similar scores, with marginal advantages of the simple model for
block validation. Interestingly, models with less basis functions in the upper partitioning levels here tend to perform slightly better.
Figure \ref{fig:fig_sst_prediction} shows mean and variance predictions of
the nonstationary model with $r = 32$ at three different days. Compared to the stationary models, 
uncertainty estimates of the nonstationary model tend to be higher, with strongest differences at the northern latitudes. Figure \ref{fig:sst_M3pars} furthermore illustrates the spatially varying standard deviation and anisotropy processes. For different values of $r$, the model shows rather similar curves.

Computation times of predictions vary considerably with model complexity. As such, random validation with the nonstationary models containing 30 parameters in the covariance function took approximately 5 to 7 times longer than with the simple model. However, these differences become smaller as $r$ increases and also for block validation, where computations with the nonstationary models were even faster. For parameter estimation, the number of needed MRA likelihood evaluations generally varies for different initial values, $M$, $r$, and was also larger for the nonstationary model with more parameters. As a result, computation times of parameter estimation are more difficult to estimate although one may set limits on convergence criteria.


\subsection{Daily Precipitation Measurements}
\label{daily-precipitation-measurements}

As a more difficult scenario, we apply the same models to precipitation data from integrated
multi-satellite retrievals (IMERG) \citep{huffman2015nasa} of the Global Precipitation Measurement Mission (GPM) \citep{hou2014global}.
The used dataset contains daily accumulated liquid precipitation recorded between 2018-08-01 and
2018-08-30 at latitudes in the range $[60^{\circ} S,60^{\circ} N]$. As in the sea surface temperature example, we average pixels to a size of $1^{\circ} \times 1^{\circ}$ to keep reasonable computation times. The dataset contains $360 \times 120 \times 30 = 1.296e6$ measurements.

Daily precipitation is highly non-Gaussian, i.e., nonnegative and zero-inflated, which is challenging for prediction based on Gaussian processes \citep[see approaches in][]{beek1992spatial, Kleiber2012}. This example here does not try to overcome these issues but is rather an optimistic, somewhat naive, practical test of whether more complex covariance models may improve prediction of such data. As a first step, we apply a Box-Cox power transform with parameters $\lambda_1 = 0.03$ and $\lambda_2 = 0.001$ and afterwards subtract transformed values from the mean. We then again select 100000  observations randomly, which we use for model fitting and predicting other observations (\emph{random validation}). However, we apply a different block validation here where we simply alternately leave out the left and right halves of images. We run each model again with different numbers of basis functions and evaluate prediction scores and computation times as in the SST example.  

\begin{table}[t]
\caption[Results of predictions in the precipitation data example]{\label{tab:precipitation_results}Results of predictions in the precipitation data example.}
\centering
\scriptsize
\setlength{\tabcolsep}{0.8mm}
\begin{tabular}{lllrrrrrrrr}
\toprule
\multicolumn{3}{c}{ } & \multicolumn{4}{c}{Random Validation} & \multicolumn{4}{c}{Block Validation} \\
\cmidrule(l{3pt}r{3pt}){4-7} \cmidrule(l{3pt}r{3pt}){8-11}
Model & M & r & RMSE & MAE & COV2SD & $t$ (min)  & RMSE & MAE & COV2SD & $t$ (min)\\
\midrule
M1 & 4 & 16 & 2.287 & 1.686 & 0.939 & \textbf{15} & 3.104 & 2.433 & 0.946 & \textbf{32}\\
M1 & 4 & 32 & 2.289 & 1.688 & 0.940 & 19 & 3.087 & 2.433 & \textbf{0.953} & 38\\
M1 & 4 & 64 & 2.278 & 1.678 & 0.939 & 30 & 3.125 & 2.427 & 0.943 & 46\\
\addlinespace[0.2em]
M2 & 4 & 16 & 2.271 & 1.678 & 0.943 & 24 & 3.116 & 2.427 & 0.941 & 40\\
M2 & 4 & 32 & 2.272 & 1.678 & 0.942 & 28 & 3.108 & 2.419 & 0.943 & 45\\
M2 & 4 & 64 & \textbf{2.263} & \textbf{1.673} & 0.941 & 41 & \textbf{3.101} & 2.411 & 0.939 & 55\\
\addlinespace[0.2em]
M3 & 4 & 16 & 2.614 & 2.025 & 0.942 & 136 & 3.116 & \textbf{2.403} & 0.837 & 131\\
M3 & 4 & 32 & 2.307 & 1.741 & \textbf{0.949} & 149 & 3.123 & 2.444 & 0.894 & 141\\
M3 & 4 & 64 & 2.265 & 1.686 & 0.945 & 164 & 3.102 & 2.434 & 0.921 & 162\\
\bottomrule
\end{tabular}
\end{table}

\begin{figure}
	\centering 
	\includegraphics[width=11cm]{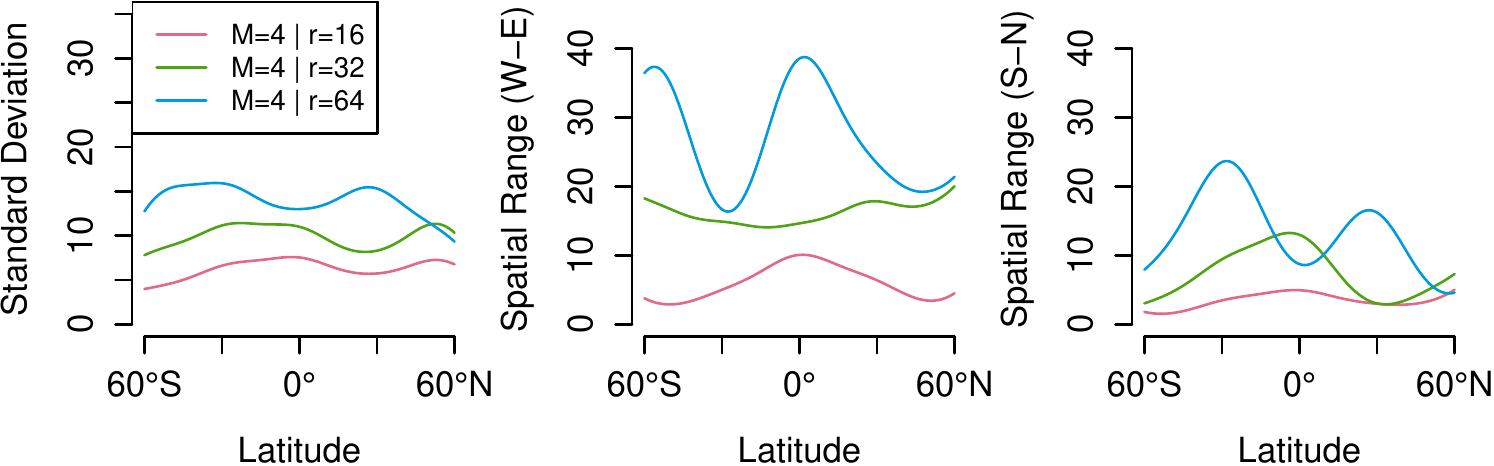} 
\caption[Estimates of spatially varying standard deviation and anisotropy processes of model \texttt{M3}  in the precipitation data example]{\label{fig:fig_prec_parameters}Estimates of spatially varying standard deviation and anisotropy processes of model \texttt{M3}  in the precipitation data example.}\label{fig:precipitation_m3pars}
\end{figure}

Table \ref{tab:precipitation_results} shows the overall results. As expected from the non-Gaussian nature of the data, prediction scores tend to be rather poor. In general, the stationary nonseparable model on spherical distances with $r=64$ tends to perform best, although for block validation, \texttt{M1} with $r=32$ provides slightly better uncertainty estimates and \texttt{M3} with $r=16$ has a slightly better MAE score. Noticeably, random validation performance of \texttt{M3} changes strongly with the number of basis functions. 
At $r=64$, results tend be similar to the best performing model, whereas at $r=16$ scores are weakest. This can be explained by stronger differences in the estimation of the spatially varying standard deviation and anisotropy processes (see Figure \ref{fig:precipitation_m3pars}). 
The computational behavior is largely in line with the sea surface temperature data and we omit a further discussion here.

\section{Discussion}
\label{discussion}

The presented experiments show that the MRA approach can be successfully applied to spatiotemporal datasets of a size prohibitive for traditional geostatistical approaches. As in \citet{katzfuss2017}, the simulation study suggests that the MRA parameters $M$ and $r$ strongly affect computation times and prediction performance and can be used to trade off computational effort against accuracy. However, no strong effect of MRA parameters on parameter estimation results has been found except that results start to become worse for higher values of $M$. The practical examples have furthermore shown that it is possible to apply the approach to global datasets, with more complex nonstationary covariance models.  Below, we discuss further results, limitations, and future research directions.

\subsection{Scalability}
\label{scalability}

The experiments showed that the order of $n \approx 10^6$ is
approachable on moderately powerful machines of individual researchers
but larger problems still require distributed computing environments. 
\citet{huang2019} provide a distributed implementation of the MRA approach
and demonstrate scalability in an example with $n > 10^7$.
Considering available global spatiotemporal datasets, 
one might still need to reduce the spatial resolution or
strongly subsample original data as in the examples. For specific practical applications, future research may
concentrate on questions such as which spatial and temporal resolution is acceptable, or whether simple models on original data or complex models on subsampled data should be preferred.

From a practical point, the most time consuming part is parameter
estimation, requiring repeated evaluations of the MRA likelihood in
numerical optimization. More parameters as in the nonstationary models
generally require more iterations. However, even for the same covariance
models, the required number of iterations varied strongly e.g. for
different numbers of basis functions used, or different initial
values, which makes computation times for parameter estimation difficult
to predict in advance.

Since the MRA approach replaces the decomposition of one huge covariance matrix by the
decomposition of many smaller matrices, future work might also try to improve
computation times on accelerated hardware such as graphics processing
units \citep{hennebohl2011}.

\subsection{Choosing MRA Parameter Values}
\label{choosing-mra-parameter-values}

The simulated experiments demonstrate that the effect of the number of
basis functions is stronger on prediction performance than on
parameter estimation (see Figures \ref{fig:fig_results_estimation_sim} and \ref{fig:fig_results_prediction_sim}). This suggests a practical strategy to select $M$ and $r$, where $M$ is 
selected first (e.g. starting with $M$ such that approximately 100 observations are available in regions on the lowest partitioning levels), using a relatively small $r$ for parameter estimation and increasing $r$ for prediction depending on available computational resources, time, and needed accuracy. The presented approach uses a rather simple regular partitioning of space and time. Since the integration of more complex spatiotemporal partitioning strategies using hierarchical discrete global grids is possible, experiments on the effect of partitioning on model efficiency would be interesting future work.


\subsection{Covariance Model Selection}
\label{covariance-model-selection}

Though the effect of different covariance models on prediction scores
was relatively small in the real world data examples, there is wide
consensus that isotropy, stationarity, and separability assumptions are
not very realistic for global scale datasets \citep[see for example][]{schmidt2020,zammitmangion2019}.

The MRA approach is very flexible in the selection of covariance
models, which can be completely defined by users. Though this opens up a 
lot of possibilities such as spatially and/or temporally varying anisotropy,
nonseparability, nonstationarity and the use of spherical distances, 
the selection of appropriate spacetime covariance models is not 
straightforward in practical applications. An existing challenge, where alternative approaches
such as the SPDE approach \citep{lindgren2011} or FRK \citep{cressie2008} have advantages
 is to combine nonstationarity with spherical covariance models. Extensions to the work in 
\citet{porcu2016} and \citet{porcu2018} adding nonstationarity are still needed.
In the present form, one may directly make use of chordal distances, the kernel 
convolution approach, and a spatiotemporal metric model to
yield a flexible class of models.

\subsection{Comparison to Other Approaches}

All in all, there are still only very few approaches that can be directly used to model spatiotemporal Gaussian processes on large global scale datasets. Compared to naive local approaches based on small neighborhoods of observations, the MRA approach may provide better predictions of large missing blocks of data and their uncertainties, and is able to infer parameters of global models, which may have physical interpretations. It combines advantages of maximum likelihood estimation and complex nonstationary spatiotemporal models, for which moments-based estimation may become impracticable. Similar to \citet{nychka15} and \citet{zammitmangion2019}, the MRA approach uses a basis function representation with generally more basis functions than available data to incorporate small and large scale spatiotemporal variations. At the same time, inference can be efficiently scaled in distributed computing environments \citep{katzfuss2017, huang2019}. The approach furthermore allows to experiment with arbitrary covariance functions, although it is still open how to add nonstationarity for spherical distances, which other approaches \citep{lindgren2011,cressie2008} include implicitly. Furthermore, an extension of the approach to handle non-Gaussian data would be needed e.g. for modeling daily precipitation. As such, integration into Bayesian hierarchical models would be possible. A systematic comparison of related spatiotemporal approaches in the style of \citet{Heaton2019} would be very helpful to identify more advantages or disadvantages, to help users with the selection of an appropriate method, and to document, how their parameters allow to trade off computation times against accuracies in parameter estimation and spatiotemporal prediction. Furthermore, first attempts to integrate deep neural networks in spatial statistics e.g. to model space deformations for nonstationarity \citep{zammitmangion2019deep}, or to learn the temporal dynamics of spatiotemporal processes with a convolutional neural network (CNN) seem very promising and may improve the modeling of complex spatiotemporal dependencies.

\section{Conclusions}
\label{conclusions}

The paper demonstrated how the multi-resolution approximation approach \citep{katzfuss2017} can be successfully used for spatiotemporal modeling of large global environmental datasets. Especially for satellite-based measurements, making use of temporal correlations can help to predict frequent larger missing spatial regions (e.g. due to cloud-cover), or to include spatiotemporal correlations in more complex models, e.g., for analyzing interactions between different environmental variables. The selection of the MRA parameters allows to effectively trade off computation times against prediction performance while at the same time allowing for flexible choice of models that include nontstationary, spatiotemporally varying anisotropy, standard deviation and smoothness. 




\section*{Code and Data Availability}
R scripts and data to reproduce the experiments are available at \url{https://github.com/appelmar/stmra_supplement}. The \texttt{stmra} R package is available at \url{https://github.com/appelmar/stmra}.

\section*{Acknowledgements}
This work was supported by the German Research Foundation (DFG) under project number 396611854.



\bibliography{references}

\appendix

\setcounter{figure}{0}    

\section{Illustration of Used Datasets}

\begin{figure}[H]
	\centering
	\includegraphics[width=11cm]{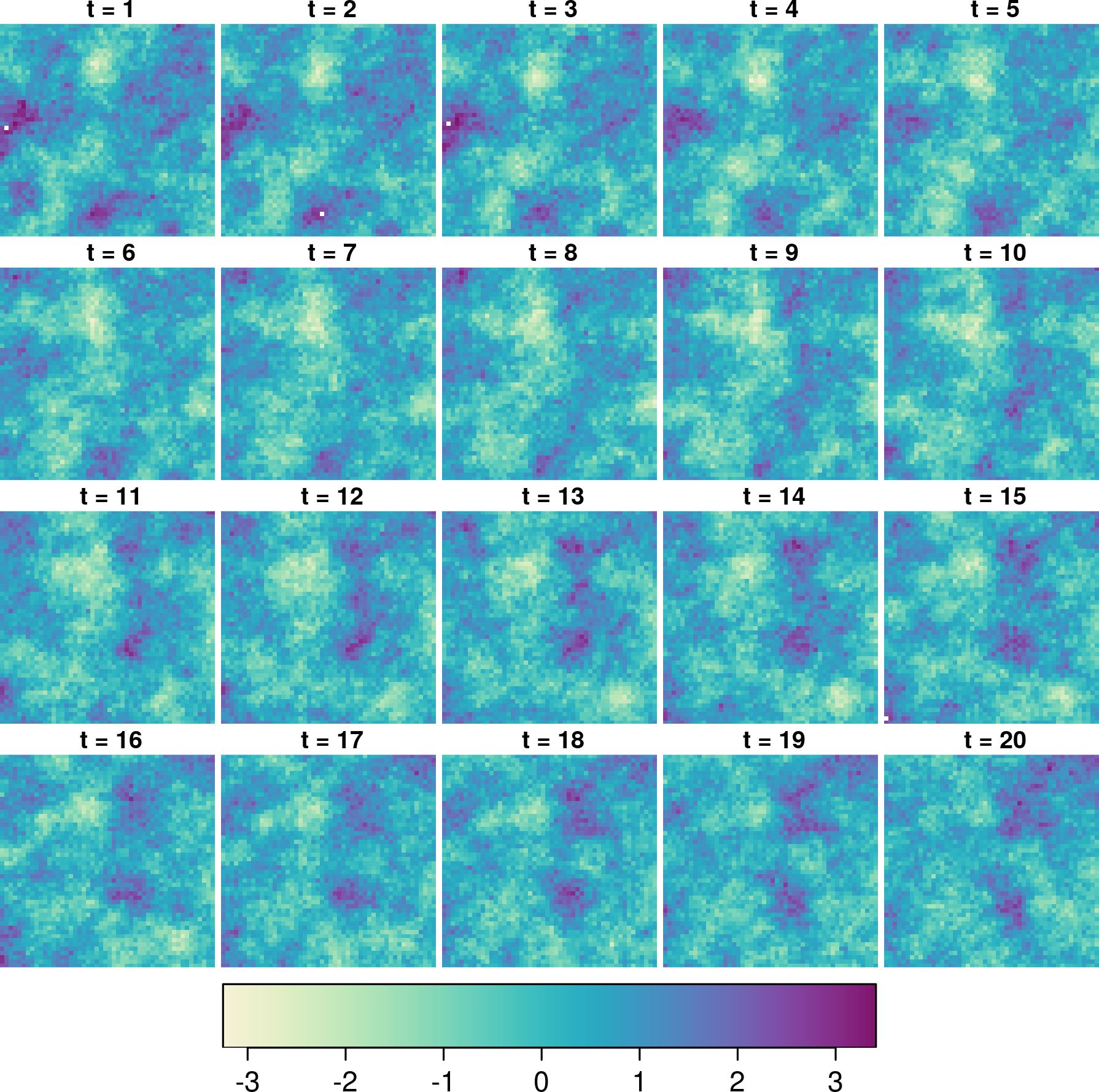}
	\caption[First 20 time slices of the simulated
		dataset]{\label{fig:simdata}First 20 time slices of the simulated
		dataset.}
\end{figure}


\begin{figure}[H]
	\centering
	\includegraphics[width=11cm]{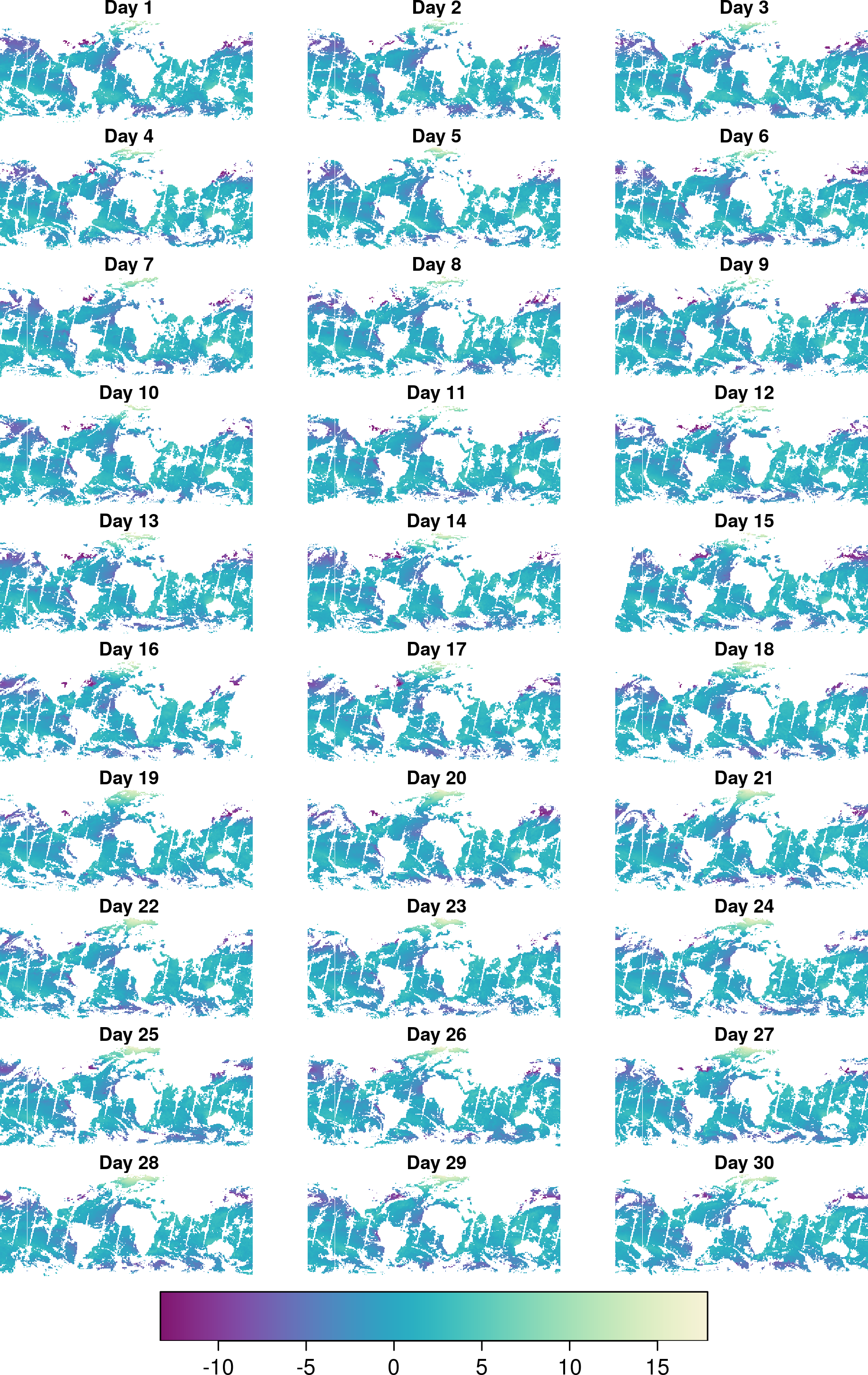}
	\caption[Daily residuals of sea surface temperatures after applying a quadratic polynomial model over latitude]{Daily residuals of sea surface temperatures after applying a quadratic polynomial model over latitude.}
	\label{fig:sst_data_all}
\end{figure}


\begin{figure}[H]
	\centering
	\includegraphics[width=11cm]{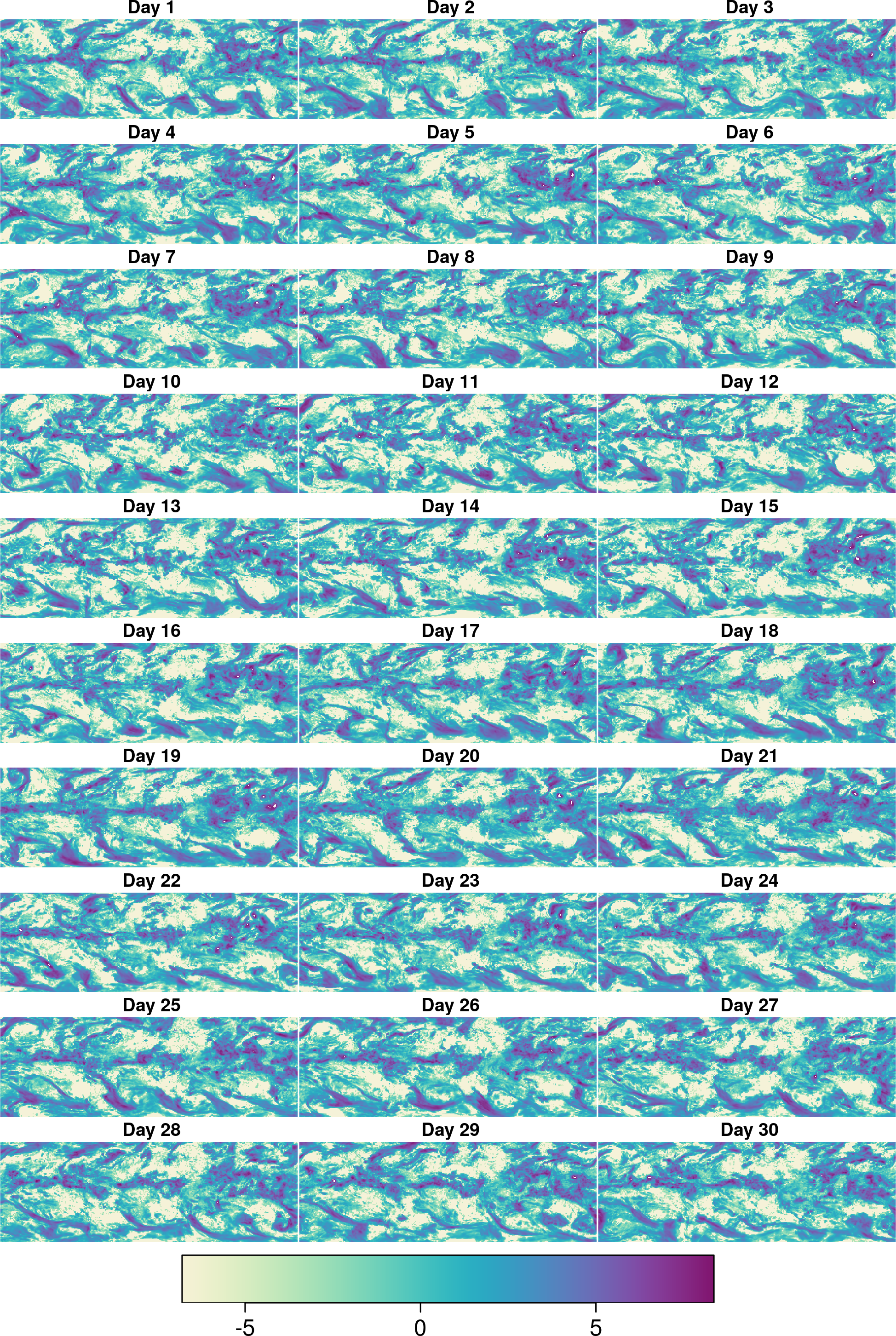}
	\caption[Daily accumulated precipitation dataset after transformation]{Daily accumulated precipitation dataset after transformation.}
	\label{fig:precipitation_data_all}
\end{figure}

\end{document}